\documentclass[12pt]{iopart}

\usepackage{graphicx} 
\begin{document}

\title{Strangeness production in hadronic models and recombination models}

\author{Gunnar Gr\"af${}^1$, Elvira Santini${}^1$, Hannah Petersen${}^{1,2}$,
Jan Steinheimer${}^1$, Michael Mitrovski${}^3$, Marcus Bleicher${}^3$}

\address{${}^1$Institut f\"ur Theoretische Physik, Goethe Universit\"at
Frankfurt,Germany} \address{${}^2$Department of Physics, Duke University,
Durham, NC 27708, United States} \address{${}^3$Frankfurt Institute for Advanced Studies
(FIAS), Ruth-Moufang-Str. 1, 60438 Frankfurt, Germany}

\ead{bleicher@th.physik.uni-frankfurt.de}

\begin{abstract} We present recent results on the production, spectra and
elliptic flow of strange particles in dynamic simulations employing hadronic
degrees of freedom and from recombination models. The main focus will be on the
Ultra-relativistic Molecular Dynamics (UrQMD) Boltzmann approach to
relativistic heavy ion collisions and a hybrid approach with intermediate
hydrodynamic evolution based on UrQMD (available for download as UrQMD v3.3).
Compared to the standard binary collision approach, an enhancement of the
strange particle particle yields is found in the hybrid approach due to the
assumption of local equilibration. The production origins of the $\phi$-meson
in the hybrid approach are studied in further detail. We also present results
on the transverse momentum spectra of baryon to meson ratios of strange
particles. Due to the approximate energy independent scaling of this ratio as a
function of $p_T$ we argue, that a maximum in these spectra may not be a unique
sign for quark coalescence but can be understood in terms of flow and
fragmentation.  \end{abstract} \maketitle

%%%%%%%%%%%%%%%%%%%%%%%%%%%%%%%%%%%%%%%%%%%%%%%%%%%%%%%%%%%%%%%%%%%%%%%%%
\section{Introduction}

Heavy ion collisions offer the unique possibility to explore the properties of
nuclear matter under extreme conditions in the laboratory. To achieve high
temperatures and particle densities, accelerators like the Relativistic Heavy
Ion Collider (RHIC) and Large Hadron Collider (LHC) or the planned FAIR facility
are used to collide heavy nuclei and to detect the remnants of their
interaction. During the collision of these nuclei it is expected that a hot and
dense zone of free quarks and gluons, known as the quark gluon plasma (QGP) is
created. 

Many years ago, strangeness has been proposed as a good tool to investigate this
new state of matter \cite{Koch:1986ud}, because the strange ($s$) and
anti-strange ($\bar{s}$) quarks are not present in the colliding nuclei and have
to be produced during the collision. An additional advantage of strange quarks
is that they decay by weak interactions, which means that they have a lifetime
on the order of $10^{-10}$ seconds which is much longer than the timescale of a
heavy ion collision and allows for a direct reconstruction of strange hadrons in
the detectors. 

On the theoretical side, the production of multi-strange (anti-)baryons and
$\phi$-mesons is still a difficult task.   Hadronic transport approaches based
on binary particle scatterings (i.e. without a QGP transition or additional
effects like strong color fields \cite{Bleicher:2000us} or multi-particle
interactions \cite{Greiner:2000tu,Cassing:2001ds}) are usually not able to
reproduce the multiplicities of multi-strange hadrons. This observation is
related to the long equilibration times and high mass threshold in the hadronic
environment compared to a QGP state.

Hydrodynamics has been suggested as an alternative tool to describe the hot and
dense stage of heavy ion collisions
\cite{Csernai:1982zz,Rischke:1995ir,Rischke:1995mt,Rischke:1996nq,Hirano:2001eu,
Kolb:2003dz,Huovinen:2008te}. Here one assumes local equilibration of the
QCD-matter, however, one has to deal with complications concerning the
decoupling stage and potential off-equilibrium effects, like viscosities or
heat conductivity. In these proceedings we present a novel integrated hybrid
approach that dynamically links a non-equilibrium Boltzmann dynamics approach
for the initial and final stage of the reaction to an intermediate ideal
hydrodynamic stage. This integrated hybrid approach is called UrQMD v3.3
\cite{urqmd:link} and is based on the known Ultra-relativistic Quantum
Molecular Dynamics. For a detailed description of the model and comparison to
data the reader is referred to
\cite{Petersen:2008dd,Petersen:2009zi,Petersen:2009vx,Petersen:2009mz}.
\begin{figure} \center
\includegraphics[width=.5\textwidth]{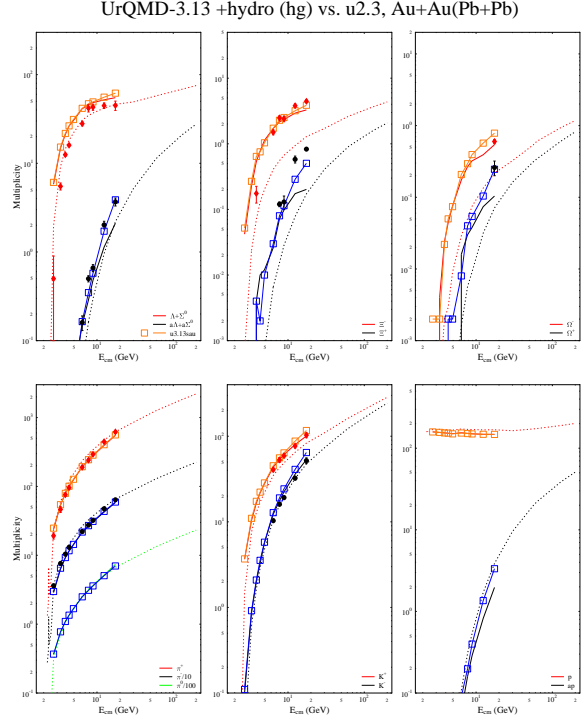}
\caption{Excitation functions of the $\Lambda$, $\Sigma$, $\Omega$, $\pi$, $K$
and $p$ multiplicities ($4\pi$) in central ($b<3.f$ fm) Au+Au/Pb+Pb collisions.
The results for the hybrid model with isochronous freeze-out (full lines), the
hybrid model with gradual freeze-out (squares) and pure UrQMD-2.3 (dotted
lines) are compared to experimental data from various experiments.
\label{fig:exc}} \end{figure}

%%%%%%%%%%%%%%%%%%%%%%%%%%%%%%%%%%%%%%%%%%%%%%%%%%%%%%%%%%%%%%%%%%%%%%%%%%

\section{Hybrid approaches} Hybrid approaches to unite hydrodynamics and
transport equations where proposed 10 years ago
\cite{Bass:1999tu,Dumitru:1999sf} and have since then been employed for a wide
range of investigations
\cite{Teaney:2001av,Socolowski:2004hw,Nonaka:2005aj,Hirano:2005wx}. The hybrid
approach presented here is based on the integration of a hydrodynamic evolution
into the UrQMD approach \cite{Petersen:2008kb}. During the first phase of the
evolution the particles are described by UrQMD as a purely hadronic cascade.
Once the two colliding nuclei have passed through each other the hydrodynamic
evolution starts at the time $t_{start}=2R/\sqrt{\gamma^2-1}$. At this time the
spectators continue to propagate in the cascade, while all other particles are
mapped to the hydrodynamic grid. By doing so one explicitly assumes a local
thermal equilibrium for each cell. The hydrodynamic evolution is performed using
the SHASTA \cite{Rischke:1995ir} algorithm with a hadron gas equation of state.
At the end of this phase the hydrodynamic fields are mapped to particle degrees
of freedom using the Cooper-Frye equation. Two possible freeze-out criteria
procedures where tested. The first is the isochronous freeze-out (IF). In this
approach, all hydrodynamic cells are mapped onto particles at the same time,
once the energy drops below five times the ground state energy density in all
cells. The second approach is called gradual freeze-out (GF). In this approach
the freeze-out happens simultaneously in a transverse slice of thickness 0.2 fm
once all cells in this slice have an energy density lower than five times the
ground state energy. This mimics an iso-eigentime freeze-out that accounts for
time dilation effects seen in the freeze-out temperature in cells with a high
rapidity. After this the final state interactions and decays of the particles
happen within the UrQMD framework.

In Fig. \ref{fig:exc} the excitation functions of the $\Lambda$, $\Sigma$,
$\Omega$, $\pi$, $K$ and $p$ yields are shown. One observes an enhancement of the
multiplicities for all particles with strange content compared to the non-hybrid
results. This enhancement leads to an improved description of the experimental
data. This is especially true for the gradual freeze-out, while the isochronous
freeze-out suffers from relativistic effects at higher beam energies. For
further discussion on spectra, the reader is referred to \cite{Petersen:2008dd}.

%%%%%%%%%%%%%%%%%%%%%%%%%%%%%%%%%%%%%%%%%%%%%%%%%%%%%%%%%%%%%%%%%%%%%%%%%%%%

\section{$\phi$-production} 
\begin{figure}
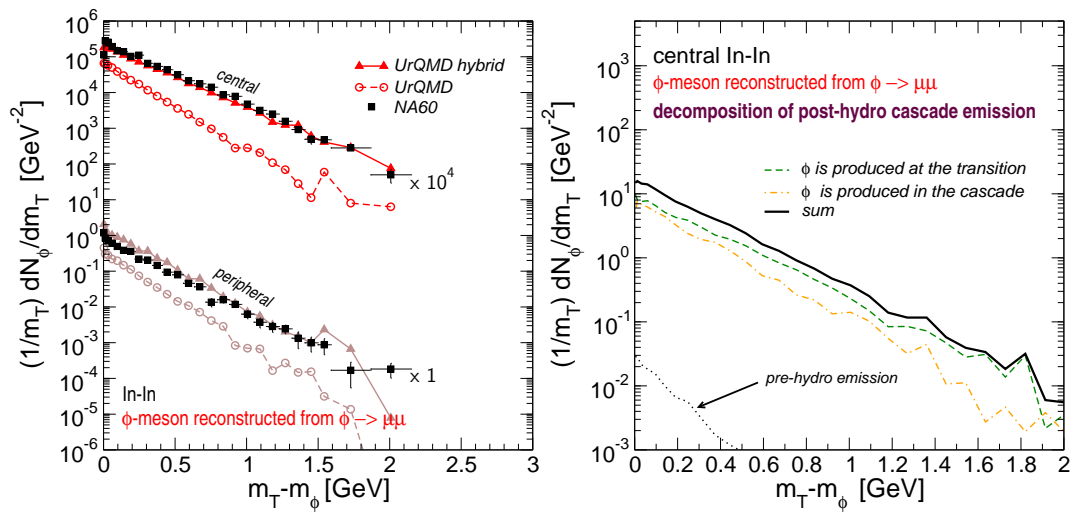
 \center
\includegraphics[width=.45\textwidth]{comp_urqmd-hyd.eps}%
\includegraphics[width=.45\textwidth]{cascade_yield_c1.eps} \caption{Left:
Transverse mass distributions of the $\phi$ meson in central (top) and
peripheral (bottom) indium-indium collisions. The hybrid model calculation
(full line) is compared to the pure UrQMD transport calculation (dashed line)
and to experimental data \cite{Arnaldi:2009wr}. Bin-widths which coincide with
the ones of the experimental data have been used. Right: Decomposition of the
post-equilibrium $\phi\rightarrow \mu^+\mu^- $ production in: (i) emission from
$\phi$ mesons which are merged into the cascade at the transition point via the
Cooper-Frye equation and emit in the cascade stage of the evolution (dashed
line); (ii) emission from $\phi$ mesons which are produced and emit in the
cascade stage (dotted-dashed line). The pre-equilibrium emission is denoted by
the dotted line.  The full line represents the total cascade emission. Due to
the smallness of the pre-equilibrium emission the latter practically coincides
with the total post-equilibrium emission.\label{compare}} \end{figure}

Apart from multi-strange baryons, $\phi$-mesons
pose a second problem in many conventional hadronic approaches. This difficulty
is not restricted to ultra-relativistic collision energies, but is also present
at lower reaction energies, see e.g. \cite{Barz:2000zz}. Recently, the present
hybrid approach has been used to investigate $\phi\rightarrow\mu^+\mu^-$
production in In+In collisions at  $E_{\rm lab}$=158$AGeV$
\cite{Santini:2009nd}. It was found that the hybrid model naturally provides a
realistic description of the $\phi$ yields and their transverse mass spectra.
In Fig. \ref{compare} (left), the hybrid model calculations are compared to
pure cascade calculations, in which no assumption of an intermediated locally
equilibrated phase is made, and to NA60 data \cite{Arnaldi:2009wr}. The two
extreme cases of central and peripheral collisions are considered. We observe
that differences in the transverse dynamics resulting from the two approaches
become stronger especially for central collisions, with NA60 data favouring the
hybrid approach to the pure transport calculation. Small deviations between the
hybrid approach calculations and the experimental data are found for peripheral
collisions where this kind of hybrid models seem to reach their  limit of
applicability \cite{Petersen:2009zi}. The transverse dynamics resulting from
pure transport calculations is comparable to the experimental data for the
peripheral reaction, though the yield remains underestimated. In this respect,
however, it should be mentioned that the UrQMD model underestimates $\phi$
meson production at high energies already in $p$+$p$ collisions. This might
partially explain that the main deviations between the experimental data and
the transport calculations for peripheral reactions are essentially related to
the yield.
%%%%%%%%%%%%%%%%%%%%%%%%%%%%%%%%%%
%%%%%%%%%%%%%%%%%%%%%%%%%%%%%%%%%%

Let us now discuss more in detail the hybrid model calculations and, in
particular, the emission during the cascade stage.  In Fig. \ref{compare}
(right) a decomposition of the various contributions from the cascade emission
is shown.  This stage contains pre- and post-equilibrium emission. The
pre-equilibrium di-muon emission is negligible, as expected from the shortness
of this stage.  The post-equilibrium emission can be divided in two categories:
(i) the emission from $\phi$ particles produced via Cooper-Frye at the
transition point and (ii) the emission from $\phi$ particles produced
\emph{during} the cascade. 

In the first case, the particles have a momentum distribution that reflects the
thermal properties of the transition point, although their di-muon decay occurs
later in a non-thermal environment.  In this sense, this copious emission,
though not specifically thermal (i.e. not described by thermal rate equations)
still carries information about the preceding thermal phase. 

The second contribution, on the contrary, can be labeled as a ``purely cascade''
one. This is the contribution of $\phi$ particles produced in the
non-equilibrium environment on the way to final decoupling. This second
contribution is characterised by steeper transverse mass spectra. the shape of
the total spectra is found to be composed by the interplay of both emissions.
Emission during the hydrodynamical phase of the evolution has been investigated
too and was found to be smaller than the post-cascade emission
\cite{Santini:2009nd}.

The analysis supports the picture of a $\phi$ meson being emitted from an
equilibrium source.  The mechanism behind the thermalization of the $\phi$ meson
cannot be inferred by this analysis and remains an open question. The possible
mechanism for the (eventual) thermalization of the $\phi$ in medium are still
under debate.  In transport calculations the dynamics of the $\phi$ meson is
typically characterised by small cross sections and early transport calculations
\cite{Ko:1993id, Haglin:1994xu} supported the idea of a large $\phi$ mean free
path in the hot hadronic fireball (see also \cite{vanHecke:1998yu}).  In the
UrQMD model, more specifically, the $\phi$ meson couples mainly to $K\bar{K}$
and $\rho\pi$ channels and its elastic and inelastic scattering with mesons and
baryons are modelled within the Additive Quark Model. Values for the cross
sections for such reactions can be found in Table IX of Ref.
\cite{Bleicher:1999xi}.  Over the years, however, some mechanisms have been
suggested that could be responsible for a stronger $\phi$ coupling to the high
density high temperature phase than expected.  Alvarez-Ruso and Koch, for
example, studied the interactions of the $\phi$ meson with other pseudo-scalar
and vector mesons in a hadronic gas at finite temperature
\cite{AlvarezRuso:2002ib}.  Within the Hidden Local Symmetry Lagrangian they
calculate the $\phi$ collision rate and mean free path including an extended
family of mechanisms that are allowed by symmetries but are not present in
calculations that rely on observed decays. In such a scenario a significantly
smaller mean free path (between 2.4 and 1 fm at temperatures above $T$=170 MeV)
was found.  Recently, the hypothesis of a catalytic $\phi$ meson production in
heavy-ion collisions has been suggested \cite{Kolomeitsev:2009yn}. When
implemented within a schematic fireball evolution scenario, such an hypothesis
can account for the increase of the $\phi$ rapidity distribution width with the
collision energy.  In a recent calculation of spectral density of the $\phi$
meson in a hot bath of nucleons and pions \cite{Vujanovic:2009wr} a significant
increase of the width of the $\phi$ meson in medium, which again supports the
scenario of a $\phi$ not at all weakly coupled to the medium. Another
possibility within the hadronic scenario might be multi-particle interaction
that could as well contribute to increase the degree of thermalization and lead
to thermalized $\phi$ production \cite{BraunMunzinger:2003zz}.
%%%%%%%%%%%%%%%%%%%%%%%%%%%%%%%%%%%%%%%%%%%%%%%%%%%%%%%%%%%%%%%%%%%
\begin{figure} \center
\includegraphics[width=.48\textwidth]{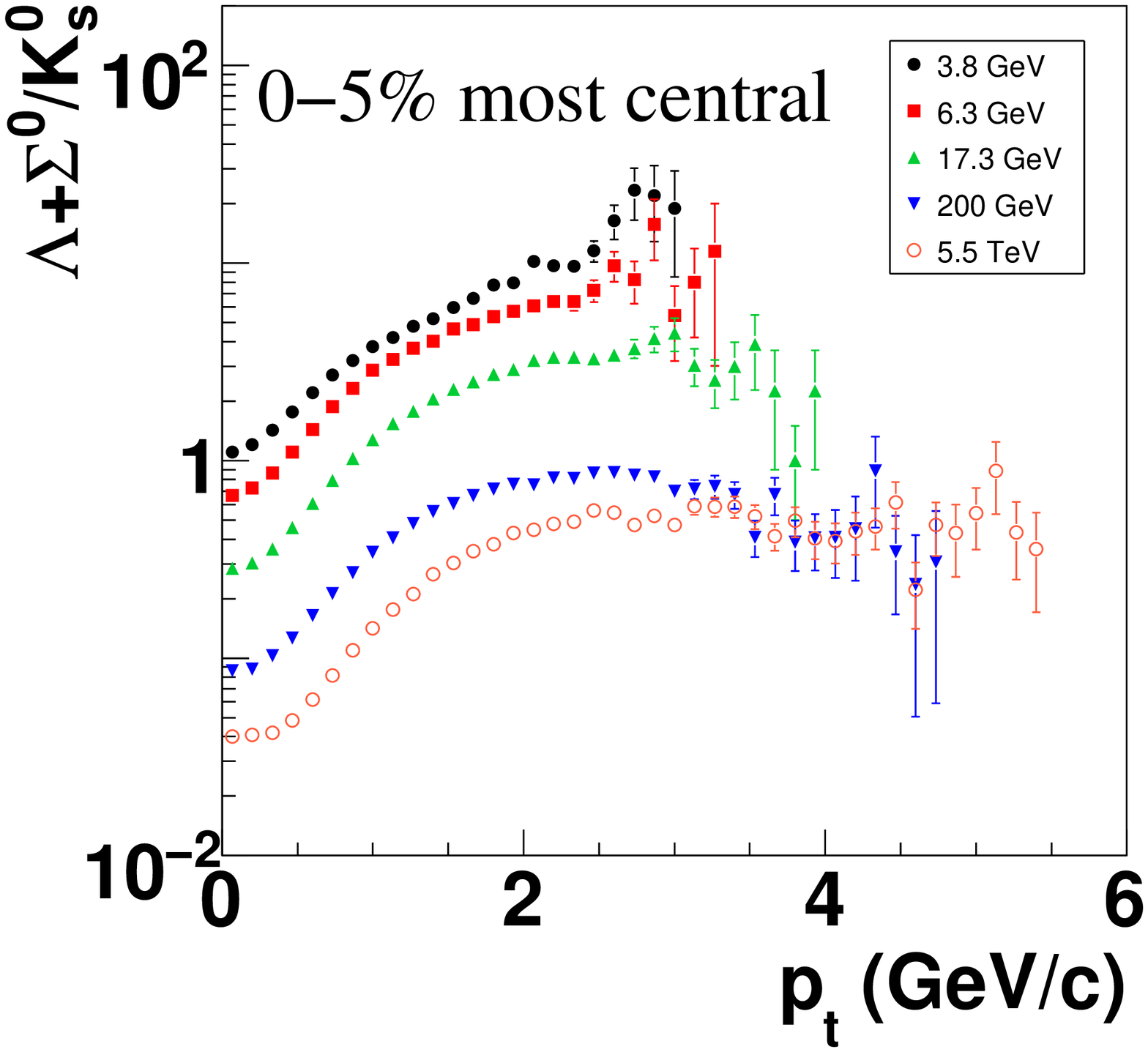}
\includegraphics[width=.48\textwidth]{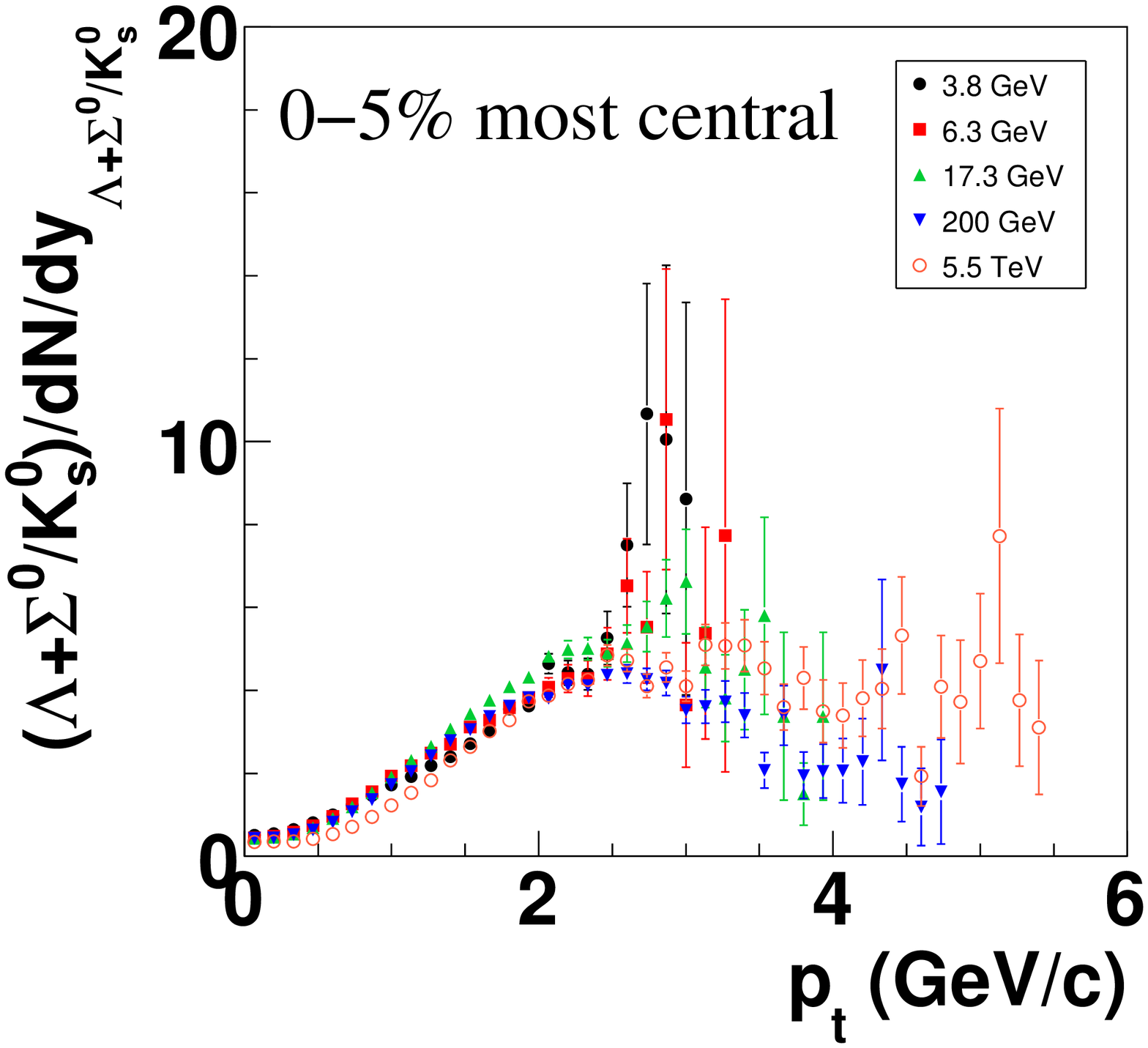}
\caption{Left: Transverse momentum distribution at midrapidity $|y|\le 0.5$ of
the $(\Lambda+\Sigma^0)/K^0_s$ ratio from $\sqrt{s}=3.8$ GeV (top) to
$\sqrt{s}=5.5$ TeV (bottom). The values were calculated with pure UrQMD-2.3 for
0-5\% most central Au+Au collisions. Right: Transverse momentum distribution at
midrapidity of the $(\Lambda+\Sigma^0)/K^0_s$ ratio scaled with the yield at
midrapidity $|y|\le 0.5$ for center of mass energies from 3.8 GeV to 5.5 TeV.
\label{fig:lambda_k0s}} \end{figure}

%%%%%%%%%%%%%%%%%%%%%%%%%%%%%%%%%%%%%%%%%%%%%%%%%%%%%%%%%%%%%%%%%%

\section{Baryon to meson ratios} At the Relativistic Heavy Ion Collider (RHIC)
at Brookhaven an unexpected enhancement of the baryon to meson ratios in central
Au+Au reactions at $\sqrt{s}=200 \rm{GeV}$ in the transverse momentum region
about 2 GeV/c had been observed. Fries et al. looked in to the baryon to meson
ratios \cite{Fries:2003kq} using a combined recombination and fragmentation
picture. Comparing their results with statistical models and data they suggest
that the maximum in the $p_\perp$ dependence of p/$\pi^0$ and $\Lambda/K^0$
ratios is a strong sign for a transition of a recombination dominated regime to
a fragmentation dominated regime at higher $p_\perp$. 

Let us investigate this effect in Fig. \ref{fig:lambda_k0s}(left). Here, the
results of pure UrQMD calculations without a hydrodynamic phase are shown. One
clearly observes a maximum of the baryon to meson ratio at about 2 GeV/c.
However, it seems that the magnitude of the enhancement is less pronounced than
in the data. To test, if the enhancement is unique to the high $p_\perp$ tail or
if it results from an overall normalisation, we normalise the $p_\perp$ spectra
to their integral.

The normalised  $(\Lambda+\Sigma^0)/K^0_s$ ratios are shown in  Fig.
\ref{fig:lambda_k0s} (right). The surprising result is a collapse of all
$(\Lambda+\Sigma^0)/K^0_s$ ratios at various center of mass energies from
3.8~GeV (lower AGS) to 5.5~TeV (LHC) to a single line. I.e. the apparent maximum
of the baryon to meson ratio at $p_\perp\approx$ 2~GeV is present at all
investigated energies.  We thus conclude, that the maximum in the baryon to
meson ratios, may not be a unique sign for quark coalescence, but rather a
natural consequence of a flow plus fragmentation picture. 

In summary, we have discussed recent developments in the area of hadron
transport simulations with emphasis on strangeness production. The most
important improvements over the last decade include the wide range employment of
hybrid approaches to simulate the hot and dense reaction stage by ideal (and
also dissipative) hydrodynamics coupled to fluctuating non-equilibrium initial
conditions and followed by a decoupling stage modelled via hadronic cascades.
These combined approaches have proven their power over a wide range of
applications ranging from strange particle production, over (elliptic) flow
analysis to leptonic and photonic probes. At the end of these proceedings we
have also shortly presented a critical view on the baryon to meson ratios and
their potential as a signal for parton recombination. 

\section*{Acknowledgements} This work was supported by the Helmholtz
International Center for FAIR within the framework of the LOEWE program launched
by the State of Hesse, GSI, BMBF and DESY. The hydrodynamic model is provided by
Dirk Rischke. The calculations have been computed at the Center for Scientific Computing
at Goethe University.\\ \vspace{1cm}

% Save this file and include it in your paper as the bibliography
% or cut and paste directly into your LaTeX

\end{document}